\documentclass{article}

\begin{document}

\title{OPERATOR PRODUCT ON LOCALLY SYMMETRIC
SPACES OF RANK ONE AND THE MULTIPLICATIVE ANOMALY}
\author{ A. A. BYTSENKO$^1$, E. ELIZALDE$^2$, M. E. X. GUIMAR\~AES$^3$ \\
\mbox{\small{1. Departamento de F\'{\i}sica, Universidade Estadual de 
Londrina}}\\
\mbox{\small{Caixa Postal 6001, Londrina--Paran\'a, Brazil}}\\
\mbox{\small{2. Institut d'Estudis Espacials de Catalunya,}}\\
\mbox{\small{Consejo Superior de Investigaciones Cientificas (IEEC/CSIC)}}\\
\mbox{\small{Edifici Nexus, Gran Capit{\`a} 2--4, 08034 Barcelona, Spain;}}\\
\mbox{\small{Departament d'Estructura i Constituents de la Mat{\`e}ria, 
Facultat de F{\`i}sica,}}\\
\mbox{\small{Universitat de Barcelona, Av. Diagonal 647, 08028 Barcelona, 
Spain}} \\
\mbox{\small{3. Universidade de Bras\'{\i}lia, Departamento de Matem{\'a}tica}}\\
\mbox{\small{CEP: 70910--900, Bras\'{\i}lia--DF, Brazil}}}

\maketitle
\begin{abstract}

The global multiplicative properties of Laplace type operators
acting on irreducible rank one symmetric spaces are considered.
The explicit form of the multiplicative anomaly is derived and its
corresponding value is calculated exactly, for important classes
of locally symmetric spaces and different dimensions.

\end{abstract}

\section{Introduction}

In theories of quantum fields (for example, in higher-derivative
quantum gravity) one has to deal with the product of two (or more)
elliptic differential operators. It is natural, therefore, to
investigate multiplicative properties of the determinants of
differential operators, in particular the so--called
multiplicative anomaly~\cite{kont94u-56,kont94u-40} (for the
definition of this anomaly see Sect. 3 below). The multiplicative
anomaly can be expressed by means of the non--commutative residue
associated with a classical pseudo--differential operator, the
Wodzicki residue~\cite{wodz87}.

Recently, the important role of this residue has been recognized
in physics. The Wodzicki residue, which is the unique extension of
the Dixmier trace to the wider class of pseudo-differential
operators~\cite{conn88-117-673,kast95-166-633}, has been
considered within the non--commutative geometrical approach to the
standard model of the electroweak 
interactions~ 
\cite{conn90-18-29,conn94,conn96-53,cham96-01,cham96-56,mart96-01}.
This residue is also used to write down the Yang-Mills action
functional. The residue formulas have also been employed for
dealing with the structure of spectral functions related to
operators acting in locally symmetric spaces~\cite{bytsenko,bytsenko1},
singularity of the zeta functions~\cite{eliz96-56}, and the
commutator anomalies of current algebras~\cite{mick94-93}. Other
recent papers along these lines can be found in Refs. 15.
The purpose of the present paper is to investigate the global
multiplicative properties of invertible elliptic operators of
Laplace type acting on a non--compact symmetric space, and related
zeta functions.

\section{The Spectral Functions}

We shall be working with irreducible rank one symmetric space $X =
G/K$ of non--compact type. Thus $G$ will be connected non--compact
simple split rank one Lie group with finite center and $K\subset G$ will be
maximal compact subgroup. Let $\Gamma\subset G$ be discrete, co-compact,
torsion free subgroup.
Let $L:C^{\infty}(V(X))\mapsto C^{\infty}(V(X))$ be partial differential
operators acting on smooth sections of vector bundles $V(X)$.
Let $\chi$ be a finite--dimensional unitary representation of $\Gamma$,
let $\{\lambda_\ell\}_{\ell=0}^{\infty}$ be the set of eigenvalues of the
second--order operator of Laplace type $L =-\Delta_{\Gamma}$ acting on
smooth sections of the vector bundle over $\Gamma \backslash X$ induced
by $\chi$, and let $n_\ell(\chi)$ denote the multiplicity of $\lambda_\ell$.

We need further a suitable regularization of the determinant of a
differential operator, since the naive definition of the product
of eigenvalues gives rise to a badly divergent quantity. We make
the choice of zeta-function regularization. The zeta function
associated with the operators ${\cal L}\equiv L+b$ has the
form

\begin{equation}
\zeta(s|{\cal L})=\sum_\ell n_\ell(\chi)\{\lambda_\ell + b\}^{-s}
\mbox{,}
\end{equation}
here $b$ is arbitrary constant (endomorphism of the vector bundle
$V(X)$), called in the physical literature the potential term.
$\zeta(s|{\cal L})$ is a well--defined analytic function for
$\Re s >\mbox{dim}(X)/2$, and can be analytically continued to a meromorphic
function on the complex plane ${\cal{C}}$, regular at $s=0$.

The following representations of $X$ up to local isomorphism
can be chosen
\begin{equation}
X=\left[ \begin{array}{ll}
SO_1(n,1)/SO(n)\,\,\,\,\,\,\,\,\,\,\,\,\,\,\,\,\,\,\,\,\,\,\,\,\,\,\,\,\,
\,\,\,\,\,\,\,\,\,({\rm I}) \\
SU(n,1)/U(n)\,\,\,\,\,\,\,\,\,\,\,\,\,\,\,\,\,\,\,\,\,\,\,\,\,\,\,\,\,\,\,
\,\,\,\,\,\,\,\,\,\,\,\,({\rm II})\\
SP(n,1)/(SP(n)\otimes SP(1))\,\,\,\,\,\,\,\,\,({\rm III})\\
F_{4(-20)}/\mbox{Spin}(9)\,\,\,\,\,\,\,\,\,\,\,\,\,\,\,\,\,\,\,\,\,\,\,\,\,\,\,\,
\,\,\,\,\,\,\,\,\,\,({\rm IV})
\end{array} \right]
\mbox{,}
\end{equation}
where ${\rm dim}\,{X}=n, 2n, 4n, 16$, respectively. Then (see for
detail \cite{will97-38-796})

\begin{equation}
SO(p,q)\stackrel{\scriptsize def}{=}\left\{\mbox{g}\in GL(p+q,{\cal{ R}}\left|_
{\mbox{det} \mbox{g}=1}^{\mbox{g}^tI_{pq}\mbox{g}=I_{pq}}\right\}\right.
\mbox{,}
\end{equation}
\begin{equation}
SU(p,q)\stackrel{\scriptsize def}{=}\left\{\mbox{g}
\in GL(p+q,{\cal{C}})\left|_{\mbox{det}
\mbox{g}=1}
^{\mbox{g}^tI_{pq}\overline{\mbox{g}}=I_{pq}}\right\}\right.
\mbox{,}
\end{equation}
\begin{equation}
SP(p,q)\stackrel{\scriptsize def}{=}\left\{{\rm g}\in GL(2(p+q),
{\cal{C}})\left|_{\mbox{g}
^tK_{pq}\overline{\mbox{g}}=K_{pq}}^{\mbox{g}^tJ_{p+q}\mbox{g}
=J_{p+q}}\right\}\right.
\mbox{,}
\end{equation}
where $I_n$ is the identity matrix of order $n$ and

\begin{equation}
I_{pq}=\left( \begin{array}{ll}
-I_p\,\,\,\,\,\, 0\\
\,\,0\,\,\,\,\,\,\,\,\, I_q \\
\end{array} \right),\,\,\,
J_{n}=\left( \begin{array}{ll}
0\,\,\,\,\,\,\,\, \,\,\,\,I_n\\
-I_n\,\,\,\,\,\,\,0 \\
\end{array} \right),\,\,\,
K_{pq}=\left( \begin{array}{ll}
I_{pq}\,\,\,\,\,\, 0\\
\,\,0\,\,\,\,\,\,\,\,\, I_{pq} \\
\end{array} \right)
\mbox{.}
\end{equation}
The groups $SU(p,q), SP(p,q)$ are connected; the group $SO_1(p,q)$ is defined
as the connected component of the identity in $SO(p,q)$ while $F_{4(-20)}$ is
the unique real form of $F_4$ with Dynkin diagram
\medskip
\begin{equation}
\bigcirc - \bigcirc = \bigcirc - \bigcirc
\end{equation}
for which the character
$(\mbox{dim}{X} - \mbox{dim}K)$ assumes the value $(-20)$ (see Ref. 17).
We assume that if $G=SO(m,1)$ or $SU(q,1)$ then $m$ is even
and $q$ is odd.

The suitable Harish--Chandra--Plancherel measure is given as follows:

\begin{equation}
|C(r)|^{-2}=C_{G}\pi rP(r)\tanh \left(a(G)r\right)
\mbox{,}
\end{equation}

\begin{equation}
a(G)=\left[ \begin{array}{ll}
\pi \hspace{0.5cm}\mbox{for $G=SO_1(2n,1)$}\\
\frac{\pi}{2} \hspace{0.5cm}\mbox{for $G=SU(q,1),\,\,\,\,\,\,\,\, q$ odd}\\
\hspace{0.8cm}\mbox{or $G=SP(m,1),\,\,\,\,\,\, F_{4(-20)}$}
\end{array} \right]
\mbox{,}
\end{equation}
while the constant $C_{G}$ and the polynomials $P(r)$ (are even polynomials 
with
Miatello coefficients $a_{2\ell}$~
\cite{Miatello1,Miatello2,Miatello3,byts96-266-1,will97-38-796,bytsenko1})
are given in the Table 1.

\vspace{1.0cm}
{\bf Table 1. \,\,\, Structure of the Harish--Chandra--Plancherel 
measure}
\begin{flushleft}
\begin{tabular}{l l l}

\hline\hline
${\rm G}$ & ${\rm C}_{{\rm G}}$ & ${\rm P}(r)$ \\
\hline\hline
${\rm SO}_1(n,1),\,\,n\geq 2$ & $\left[2^{2n-4}
\Gamma\left(\frac{n}{2}\right)^2
\right]^{-1}$ &

$\prod_{k=0}^{n-2}\left[r^2+\frac{(2k+1)^2}{4}\right],\,\,
n=2m$\\
& &
$\prod_{k=0}^{n-1}\left[r^2+k^2\right],\,\,n=2m+1$ \\
${\rm SU}(n,1),\,\,n\geq 2$ & $\left[2^{2n-1}\Gamma(n)^2
\right]^{-1}$ &
$\prod_{k=1}^{n-1}\left[\frac{r^2}{4}+\frac{(n-2k)^2}{4}
\right]$\\
${\rm SP}(n,1),\,\,n\geq 2$ & $\left[2^{4n+1}\Gamma(2n)^2
\right]^{-1}$ &
$\left[\frac{r^2}{4}+\frac{1}{4}\right]\prod_{k=3}^{n+1}
\left[\frac{r^2}{4}+\left(n-k+\frac{3}{2}\right)^2\right]$\\
& &
$\times\left[\frac{r^2}{4}+\left(n-k+\frac{5}{2}\right)^2\right]$\\
${\rm F}_{4(-20)}$ & $\left[2^{21}\Gamma(8)^2
\right]^{-1}$ &
$\left[\frac{r^2}{4}+\frac{1}{4}\right]\left[\frac{r^2}{4}+\frac{9}{4}
\right]$\\
& &
$\times\prod_{k=0}^{4}
\left[\frac{r^2}{4}+\left(\frac{2k+1}{2}\right)^2\right]$
\\
\hline \hline
\end{tabular}
\end{flushleft}

\subsection{The Miatello coefficients and explicit values of anomalies}

The coefficients of the polynomial $P(r)$,
will be denoted by $a_{2\ell}$:
$$
P(r) = \sum_{\ell=0}^{n/2-1}a_{2\ell}r^{2\ell} \hspace{1.0cm}
\mbox{for}\,\,\, G\neq SO_1(2m+1,1),
$$
\begin{equation}
\,\,\,\,\,\,\,\,\,\,\,\,\,\,\,\,\,\,\,\,\,\,\,\,
\,\,\,\,
=\sum_{\ell=0}^m a_{2\ell}r^{2\ell} \hspace{1.3cm}\mbox{for}\,\,\, G=SO_1(2m+1,1).
\end{equation}
For the various rank one simple groups $G$ the Miatello coefficients
of the polynomial $P(r)$,\,
$a_{2\ell} = (1/2\ell!)[(d^{2\ell}/dr^{2\ell})P(r)]|_{r=0}$,
are given in the Table 2.

\vspace{1.0cm}
{\bf Table 2. \,\,\, First Miatello coefficients}
\begin{flushleft}
\begin{tabular}{l l l}
\hline\hline
${\rm G}$ & $ n $ & $ a_{2\ell}$ \\
\hline\hline
\\
${\rm SO}_1(n,1)$ &2 $ $ & 

$a_0 = \frac{1}{4},\, a_2 = 1$\\

& &  
$ $ \\
$$ &4 $ $ & 
$a_0=\frac{9}{16},\, a_2=\frac{5}{2},\, a_4= 1  $\\
& &
$ $\\        
$$ &6 $ $ &
$a_0=\frac{225}{64},\, a_2=\frac{259}{16},\, a_4=\frac{35}{4},\, a_6=1 $\\
& &
$ $
\\        
$$ &8 $ $ &
$a_0=\frac{11025}{256},\, a_2=\frac{3229}{16},\, a_4=\frac{987}{8},\, 
a_6 =21,\, a_8=1 $\\
& &
$ $
\\ 
$$ &10 $ $ &
$a_0=\frac{893025}{1024},\, a_2=\frac{1057221}{256},\, 
a_4=\frac{86405}{32}, \, a_6=\frac{4389}{8}, $\\
& &
$ $
\\ 
$$ & $ $ &
$a_8=\frac{165}{4},\, a_{10}=1 $\\
& &
$ $
\\

${\rm SU}(n,1)$ &2 $ $ & 

$a_0 = 0,\, a_2 = \frac{1}{4}$\\

& &  
$ $ \\
$$ &3 $ $ & 
$a_0=\frac{1}{16},\, a_2=\frac{1}{8},\, a_4= \frac{1}{16}  $\\
& &
$ $\\        
$$ &4 $ $ &
$a_0= 0,\, a_2=\frac{1}{4},\, a_4=\frac{1}{8},\, a_6=\frac{1}{64} $\\
& &
$ $
\\        
$$ &5 $ $ &
$a_0=\frac{81}{256},\, a_2=\frac{45}{64},\, a_4=\frac{59}{128},\, 
a_6 =\frac{5}{64},\, a_8=\frac{1}{256} $\\
& &
$ $
\\ 
$$ &6 $ $ &
$a_0= 0,\, a_2= 4,\, a_4=\frac{5}{2}, a_6=\frac{33}{64},\, 
a_8=\frac{5}{128}, \, a_{10}=\frac{1}{1024}$\\
& &
$ $
\\

${\rm SP}(n,1)$ &2 $ $ &


$a_0= \frac{9}{64},\, a_2=\frac{19}{64},\, a_4=\frac{11}{64},\, 
a_6=\frac{1}{64} $\\
& &
$ $
\\        
$$ &3 $ $ &
$a_0= \frac{2025}{1024},\, a_2= \frac{4581}{1024},\, a_4=\frac{1565}{512},\, 
a_6=\frac{309}{512}, $\\
& &
$ $
\\ 
$$ & $ $ &
$a_8=\frac{45}{1024},\, a_{10}=\frac{1}{1024} $\\
& &
$ $
\\

${\rm F}_{4(-20)}$ &  $ $ & 

$a_0 = \frac{8037225}{16384},\, a_2 = \frac{18445239}{16384}, \,
a_4=\frac{13020525}{16384}, \, a_6=\frac{2864323}{16384}, $\\

& &
$ $
\\        
$$ & $ $ &
$a_8= \frac{262075}{16384},\, a_{10}=\frac{10437}{16384},\, 
a_{12}=\frac{175}{16384},\, a_{14}=\frac{1}{16384} $\\
& &
$ $
\\

\hline \hline
\end{tabular}
\end{flushleft}

\newpage
\section{The Multiplicative Anomaly and Associated One--Loop Contributions}

The spectral zeta function associated with the product $\otimes {\cal L}_j$
has the form

\begin{equation}
\zeta(s|\otimes {\cal L}_j)=\sum_{\ell\geq 0}n_\ell\prod_j^2
(\lambda_\ell+b_j)^{-s}
\mbox{.}
\end{equation}
We shall always assume that $b_1\neq b_2$, say $b_1>b_2$. If $b_1=b_2$ then
$\zeta(s|\otimes{\cal L}_j)=\zeta(2s|{\cal L})$ is a well--known function.
For $b_1,b_2\in{\cal{ R}}$, set $B_{j}\stackrel{def}{=}b_j+ 1/4$.

We are interested in
multiplicative properties of determinants, the multiplicative anomaly~
\cite{wodz87,kont94u-56,kont94u-40}, associated with one--loop approximation
in quantum theory. The partition function
$\mbox{log}Z\propto-\mbox{log}\mbox{det}\left(\otimes{\cal L}_j\right)$
of the product of two elliptic differential operators for the simplest $O(2)$
invariant model of self--interacting charged fields~
\cite{bens91-44-2480} has been analyzed in Ref. 23.
The loop  approximation can be given in terms of the
multiplicative anomaly $F({\cal L}_1,{\cal L}_2)$, which has the form

\begin{equation}
F({\cal L}_1,{\cal L}_2)=\mbox{det}_\zeta[\otimes {\cal L}_j]
[\mbox{det}_\zeta({\cal L}_1)\mbox{det}_\zeta({\cal L}_2)]^{-1}
\mbox{,}
\end{equation}
where we assume a zeta--regularization of determinants, i.e.

\begin{equation}
{\rm det}_\zeta({\cal L}_j)\stackrel{def}{=}\exp\left(-\frac{\partial}
{\partial s}\zeta(s=0|{\cal L}_j)\right)
\mbox{.}
\end{equation}
Generally speaking, if the multiplicative anomaly related to elliptic operators
is nonvanishing then the relation $\mbox{log}\mbox{det}(\otimes {\cal L}_j)=
\mbox{Tr}\,\mbox{log}(\otimes {\cal L}_j)$ does not hold.

\subsection{The residue formula and the multiplicative anomaly}

The value of $F({\cal L}_1,{\cal L}_2)$ can be expressed by means of the
non--commutative Wodzicki residue~\cite{wodz87}. 
Let ${\cal O}_j$ be invertible elliptic pseudo--differential operators
of real non--zero
orders $\alpha$ and $\beta$ such that $\alpha+\beta\neq 0$. Even if the 
zeta functions
for operators ${\cal O}_1, {\cal O}_2$ and ${\cal O}_1\otimes{\cal O}_2$ are
well defined and if their principal symbols obey the Agmon--Nirenberg condition
(with appropriate spectra cuts) one has in general that

$F({\cal O}_1,{\cal O}_2)\neq 1$. For such invertible elliptic operators the
formula for the anomaly of commuting operators holds~\cite{eliz97u-60}:

\begin{equation}
{\cal A}({\cal O}_1,{\cal O}_2)={\cal A}({\cal O}_2,{\cal O}_1)=
\mbox{log}(F({\cal O}_1,{\cal O}_2))=
\frac{\mbox{res}\left[(\mbox{log}({\cal O}_1^{\beta}
\otimes{\cal O}_2^{-\alpha}))^2\right]}{2\alpha\beta(\alpha+\beta)}
\mbox{.}
\end{equation}
More general formulae have been derived in Refs. 1, 2.
%
In the case of the product of two operators the following result holds
(see Ref. 12):

{\it Theorem:}

{\it The explicit form of the multiplicative anomaly in the case
${\cal O}_j\equiv{\cal L}_j$ is
$$
{\cal A}({\cal  L}_1,{\cal
L}_2)=\frac{A}{2}\sum_{\ell=0}^{n/2-1}
\frac{(-1)^{\ell}a_{2\ell}}{(\ell+1)!}B_2^{\ell+1}
\sum_{k=1}^{\infty}\frac{\sigma_k (\ell+k+1)!}{(k+1)!}
\left(\frac{B_1-B_2}{B_1}\right)^{k+1}
$$
\begin{equation}
+\frac{A}{2}{\rm log}\left(\frac{B_1}{B_2}\right)
\sum_{\ell =0}^{n/2-1}\frac{(-1)^{\ell +1}a_{2\ell}}
{(\ell+1)}\left(B_1^{\ell+1}-B_2^{\ell+1}\right),
\end{equation}
where $A=(1/4)\chi(1){\rm Vol}(\Gamma\backslash G)C_G$ and 
$\sigma_\ell\stackrel{def}{=}\sum_{k=1}^\ell k^{-1}$.
}

Finally the numerical values of 
${ A} \equiv {\cal A}({\cal  L}_1,{\cal L}_2)
/(\chi(1){\rm Vol}(\Gamma\backslash G))$ related to the multiplicative 
anomaly ${\cal A}({\cal  L}_1,{\cal L}_2)$ for various 
rank one groups $G$ and constants $[B_1;\, B_2]$ are given in Table 3.

\vspace{0.5cm}
{\bf Table 3.\,\,\, Explicit numerical values of ${A}$}
\begin{flushleft}
\begin{tabular}{l l l}
\hline\hline
${\rm G}$ & $ n $ & $ {A}\,\,\, {\rm for\,\,\, some\,\,\,
pairs\,\,\, of}\,\,\,[B_1;B_2]$ \\
\hline\hline
\\
${\rm SO}_1(n,1)$ &4 $ $ &
$ {[1;\frac{1}{4}]}:\,\, 0.141936,$
\\
\\
$$
&
$$
&
$ {[10;\frac{1}{4}]}: \,\,\, 11.0951,$
\\
\\
$$
&
$$
&
${[200; 100 + \frac{1}{4}]}: \,\,\, 4.57424$ 
\\
\\
$$
&
$6$
&
${[1;\frac{1}{4}]}: \,\,\, -0.0636199,$
\\
\\
$$
&
$$
&
$[10;\frac{1}{4}]: \,\,\, -20.1404,$
\\
\\
$$
&
$$
&
${[200; 100 + \frac{1}{4}]}: \,\,\, -736.671$ 
\\
\\
$$
&
$8$
&
${[1;\frac{1}{4}]}: \,\,\, 0.00716077,$
\\
\\
$$
&
$$
&
${[10;\frac{1}{4}]}: \,\,\, 14.3917,$
\\
\\
$$
&
$$
&
$[200; 100 + \frac{1}{4}]: \,\,\, 25403.1$  
\\
\\ 
$$
&
$10$
&
${[1;\frac{1}{4}]}: \,\,\, -0.000551611,$
\\
\\
$$
&
$$
&
${[10;\frac{1}{4}]}: \,\,\, -6.66525,$
\\
\\
$$
&
$$
&
${[200; 100 + \frac{1}{4}]}: \,\,\, -599988.$ 
\\ 
\\ 
\end{tabular}
\end{flushleft}

\newpage

\begin{flushleft}
\begin{tabular}{l l l}
\hline\hline
${\rm G}$ & $ n $ & $ { A}\,\,\, {\rm for\,\,\, some\,\,\,
pairs\,\,\, of}\,\,\,[B_1;B_2]$ \\
\hline\hline
\\
${\rm SU}(n,1)$ &4 $ $ &
${[1;\frac{1}{4}]}: \,\,\, 0.00110888,$
\\
\\
$$
&
$$
&
${[10;\frac{1}{4}]}: \,\,\, 0.0866804,$
\\
\\
$$
&
$$
&
${[200; 100 + \frac{1}{4}]}: \,\,\, 0.0357362$ 
\\
\\ 
$$
&
$6$
&
${[1;\frac{1}{4}]}: \,\,\, -0.000150991,$
\\
\\
$$
&
$$
&
${[10;\frac{1}{4}]}: \,\,\, -0.0488322,$
\\
\\
$$
&
$$
&
${[200; 100 + \frac{1}{4}]}: \,\,\, -1.79837$  
\\
\\
$$
&
$8$
&
$[1;\frac{1}{4}]: \,\,\,  5.82224\times 10^{-6},$
\\
\\
$$
&
$$
&
$[10;\frac{1}{4}]: \,\,\, 0.0103493,$
\\
\\
$$
&
$$
&
$[200; 100 + \frac{1}{4}]: \,\,\, 17.8461$  
\\
\\
$$
&
$10$
&
$[1;\frac{1}{4}]: \,\,\, -1.13623\times 10^{-7},$
\\
\\
$$
&
$$
&
$[10;\frac{1}{4}]: \,\,\, -0.00102054,$
\\
\\
$$
&
$$
&
$[200; 100 + \frac{1}{4}]: \,\,\, -85.8268$ \\ 
\\
\\
${\rm SP}(n,1)$ &4 $ $ &
$[1;\frac{1}{4}]: \,\,\, 0.0000152281,$
\\
\\
$$
&
$$
&
$[10;\frac{1}{4}]: \,\,\, 0.00119037,$
\\
\\
$$
&
$$
&
$[200; 100 + \frac{1}{4}]: \,\,\, 0.000490762$  
\\
\\
$$
&
$6$
&
$[1;\frac{1}{4}]: \,\,\, -2.63879\times 10^{-8},$
\\
\\
$$
&
$$
&
$[10;\frac{1}{4}]: \,\,\, -8.51064\times 10^{-6},$
\\
\\
$$
&
$$
&
$[200; 100 + \frac{1}{4}]: \,\,\, -0.000313153$ 
\\
\\

\end{tabular}
\end{flushleft}

\newpage

\begin{flushleft}
\begin{tabular}{l l l}
\hline\hline
${\rm G}$ & $ n $ & $ {A}\,\,\, {\rm for\,\,\, some\,\,\,
pairs\,\,\, of}\,\,\,[B_1;B_2]$ \\
\hline\hline
\\
$$
&
$8$
&
$[1;\frac{1}{4}]: \,\,\, 8.70798\times 10^{-12},$
\\
\\
$$
&
$$
&
$[10;\frac{1}{4}]: \,\,\,  1.5324\times 10^{-8},$
\\
\\
$$
&
$$
&
$[200; 100 + \frac{1}{4}]: \,\,\, 0.0000263649$ 
\\
\\
$$
& 
$10$
&
$[1;\frac{1}{4}]: \,\,\, - 1.30145\times 10^{-15},$
\\
\\
$$
&
$$
&
$[10;\frac{1}{4}]: \,\,\, -1.13401\times 10^{-11},$
\\
\\
$$
&
$$
&
$[200; 100 + \frac{1}{4}]: \,\,\, -9.46242\times 10^{-7}$ 
\\
\\

${\rm F}_{4(-20)}$ & $ $ &
$[1;\frac{1}{4}]: \,\,\, -2.09552\times 10^{-11},$
\\
\\
$$
&
$$
&
$[10;\frac{1}{4}]: \,\,\, -6.76666\times 10^{-9},$
\\
\\
$$
&
$$
&
$[200; 100 + \frac{1}{4}]: \,\,\, -2.49078\times 10^{-7}$ \\ 
\\
\\

\hline \hline
\end{tabular}
\end{flushleft}

\section{Conclusions}

In this paper, the multiplicative properties of operators of Laplace
type and their related zeta functions have been studied.
Explicit formulas for the multiplicative anomaly have been investigated.
In a special case, namely for $n=2$, Theorem 1 gives ${\cal A}({\cal L}_1,
{\cal L}_2)=0$. For any odd $n$, the multiplicative anomaly is zero.
This statement follows from the general theory of Laplace--type operators 
(see, for example, Ref. 23).
Note that for the four--dimensional space with $G=SO_1(4,1)$, one derives
from Theorem 1 the result
${\cal A}({\cal L}_1,{\cal L}_2)=-A_G(b_1-b_2)^2,\, n=4$,
which also follows from Wodzicki's formula (13), where $A_G=Aa_2/4$.

We have preferred to
limit ourselves here to discuss in detail various particular cases and
emphasize the general picture.
It seems to us that the explicit results for the
anomaly are not only interesting as mathematical results but are also
of physical interest, in view of their applications to concrete problems in
field theory and gravity, both at the classical and quantum levels.
As we see from our numerical results, some nice patterns show up. It is
interesting to notice that for all groups except $F_{4(-20)}/{\rm Spin}(9)$,
there is a dependence of the sign of the anomaly with the dimension
$n=2m$: for even $m$ the anomaly is positive while for odd $m$ it 
is negative.
Also, generically, the absolute value of the anomaly diminishes with $n$
for small values of $B_1$, $B_2$, but it can also get quite large, and 
increase,
for bigger values of $B_1$ and $B_2$.
Note also, that the spectral properties of products of differential 
operators 
related to higher-spin fields may differ, in principle, from the 
properties 
considered in this paper, what deserves further study. We hope to 
return soon 
to this problem elsewhere.


\begin{thebibliography}{10}



\bibitem{kont94u-56}
M. Kontsevich and S. Vishik, {\em "Determinants of Elliptic 
Pseudo-Differential
Operators"}, Preprint MPI/94-30 (1994).

\bibitem{kont94u-40}
M. Kontsevich and S. Vishik, {\it Geometry of Determinants of Elliptic
Operators}, e--Print arXiv hep-th/9406140.

\bibitem{wodz87}
M. Wodzicki, in {\it Lecture Notes in
Mathematics}, ed.  Yu. I. Manin (Springer--Verlag, Berlin, 1987), 
Vol. 1289, p. 320.

\bibitem{conn88-117-673}
A. Connes, {\it Commun. Math. Phys.} {\bf 117}, 673 (1988).

\bibitem{kast95-166-633}
D. Kastler, {\it Commun. Math. Phys.} {\bf 166}, 633 (1995).


\bibitem{conn90-18-29}
A. Connes and J. Lott, {\it Nucl. Phys.} {\bf B 18}, 29 (1990).

\bibitem{conn94}
A. Connes, {\it Non--Commutative Geometry} (Academic Press, New York, 1994).

\bibitem{conn96-53}
A. Connes, {\it Commun. Math. Phys.} {\bf 182}, 155 (1996).

\bibitem{cham96-56}
A. H. Chamseddin and A. Connes, {\it Phys. Rev. Lett.} {\bf 77}, 4868 (1996).

\bibitem{cham96-01}
A. H. Chamseddine and A. Connes, {\it Commun. Math. Phys.} {\bf 186}, 731 (1997).

\bibitem{mart96-01}
C. P. Martin, J. M. Gracia--Bondia and J. C. Varilly, {\it Phys. Reports} {\bf 294},
363 (1998).

\bibitem{bytsenko}
A. A. Bytsenko and F. L. Williams, 
{\it JMP} {\bf 39}, 1075 (1998).

\bibitem{bytsenko1}
A. A. Bytsenko and F. L. Williams, 
{\it J. Phys. A: Math. Gen.} {\bf 32}, 5773 (1999).

\bibitem{eliz96-56}
E. Elizalde, {\it J. Phys.} {\bf A 30}, 2735 (1997).

\bibitem{mick94-93}
J. Mickelsson, in   
{\it Lecture Notes in Phys.} (Springer, Berlin, 1994), Vol. 436, pp. 123--135.

\bibitem{rp1} G. Cognola, E. Elizalde  and S. Zerbini,  {\sl Dirac Functional
Determinant in Terms of the Eta Invariant and the Noncommutative
Residue}, {\it Commun. Math. Phys.}, to appear; 

E. Elizalde, {\it JHEP} {\bf
9907}, 015 (1999); 

E. Elizalde, G. Cognola and S. Zerbini, {\it Nucl.
Phys.} {\bf B532}, 407 (1998).

\bibitem{will97-38-796}
F.L. Williams, {\it J. Math. Phys.} {\bf 38}, 796 (1997).

\bibitem{Helgason}
S. Helgason, {\it Differential Geometry and Symmetric Spaces}, Pure and Applied
Math. Ser. {\bf 12}, Academic Press (1962).

\bibitem{Miatello1}
R. Miatello, {\it The Minakshisundaram--Pleijel Coefficients for the Vector--Valued
Heat Kernel on Compact Locally Symmetric Spaces of Negative Curvature}, PhD Thesis,
Rutgers University (1976).

\bibitem{Miatello2}
R. Miatello, {\it Manusckripta Math.} {\bf 29}, 249 (1979).

\bibitem{Miatello3}
R. Miatello, {\it Trans. Am. Math. Soc.} {\bf 260}, 1 (1980).

\bibitem{byts96-266-1}
A.A. Bytsenko, G. Cognola, L. Vanzo and S. Zerbini, {\it Phys. Reports} {\bf 266},
1 (1996).

\bibitem{bens91-44-2480}
K. Benson, J. Bernstein and S. Dodelson, {\it Phys. Rev.} {\bf D 44}, 2480 (1991).

\bibitem{eliz97u-60}
E. Elizalde, L. Vanzo and S. Zerbini, {\it Commun. Math. Phys.} {\bf 194},
613 (1998).




\end{thebibliography}
\end{document}